# Context Detection for Advanced Self-Aware Navigation using Smartphone Sensors

Han Gao and Paul D. Groves

*(University College London, United Kingdom)*
(E-mail: han.gao.14@ucl.ac.uk & p.groves@ucl.ac.uk)

Navigation and positioning systems dependent on both the operating environment and the behaviour of the host vehicle or user. The environment determines the type and quality of radio signals available for positioning and the behaviour can contribute additional information to the navigation solution. In order to operate across different contexts, a context-adaptive navigation solution introduces an element of self-awareness by detecting the operating context and configuring the positioning system accordingly. This paper presents the detection of both environmental and behavioural contexts as a whole, building the foundation of a context-adaptive navigation system. Behavioural contexts are classified using measurements from accelerometers, gyroscopes, magnetometers and the barometer by supervised machine learning algorithms, yielding an overall 95% classification accuracy. A connectivity dependent filter is then implemented to improve the behavioural detection results. Environmental contexts are detected from GNSS measurements. They are classified into indoor, intermediate and outdoor categories using a probabilistic support vector machine (SVM), followed by a hidden Markov model (HMM) used for time-domain filtering. As there will never be completely reliable context detection, the paper also shows how environment and behaviour association can contribute to reducing the chances of the context determination algorithms selecting an incorrect context. Finally, the proposed context-determination algorithms are tested in a series of multi-context scenarios.

1. INTRODUCTION. Navigation and positioning is inherently dependent on the context, which comprises both the operating environment and the behaviour of the host vehicle or user [1]. For many daily applications, the environment and host behaviour are subject to change, particularly for smartphones, which move between indoor and outdoor environments and can be stationary, on a pedestrian, or on any type of vehicle. To meet the growing demand for greater accuracy and reliability in a wider range of challenging contexts, many navigation and positioning techniques have been developed or improved [2], such as Wi-Fi positioning [3][4], multiple-constellation global navigation satellite system (GNSS) [5], 3D-mapping aided (3DMA) GNSS ranging [6][7] and pedestrian dead reckoning (PDR) using step detection [2][8]. However, no single current technique is able to provide reliable and accurate positioning in all contexts. Therefore, in order to operate across different contexts, a multi-sensor self-aware navigation solution is required to detect the operating context and configure the positioning system accordingly, which is also referred to as context-adaptive navigation [1][9].

    Context is critical to the operation of a navigation or positioning system. The environment determines the type and quality of radio signals for positioning. For example, GNSS reception is good in open-sky environments, but poor indoors and in deep urban areas. Wi-Fi signals are not available in rural areas, in the air or at sea. In an underwater environment, most radio signals do not propagate at all. Processing techniques can also depend on the environments. Terrain referenced navigation typically determines terrain height using radar or laser scanning in the

air, sonar or echo sounding at sea and a barometer on land [1]. In an open-sky environment, non-line-of-sight (NLOS) reception of GNSS signals or multipath interference may be detected using consistency checking techniques based on sequential elimination [10]. In dense urban areas, more sophisticated algorithms are required for GNSS positioning in the presence of severe multipath interference and NLOS reception [11].

The behaviour can contribute additional information to the navigation solution. It will help mobile devices to understand what the user is doing under particular circumstances [12]. A stationary pedestrian or a land vehicle indicates a fixed location and will not need to update its velocity and position. Land vehicles normally remain on the ground, effectively removing one dimension from the position solution. Similarly, boats, ships and underwater vehicles can all be on land, but only exhibit some specific types of behaviours. Within a GNSS receiver, the behaviour can be used to set the bandwidths of the tracking loop and coherent integration intervals, and to predict the temporal characteristics of multipath [13].

Although behavioural and environmental context can be detected separately, they are not completely independent. Certain activities are associated with certain environments in reality [1]. For example, a micro air vehicle (MAV) flies in the air indoors or outdoors, not at the bottom of the sea. All road vehicles are associated with driving, but only off-road vehicles are associated with off-road driving. A bus typically travels more slowly and stops more in cities than on the highway. A stationary pedestrian will stay in the same place during the period. This information can be used to estimate the likelihood of the detected behaviour and environment combinations and reduce the chances of the context determination algorithms selecting an incorrect context.

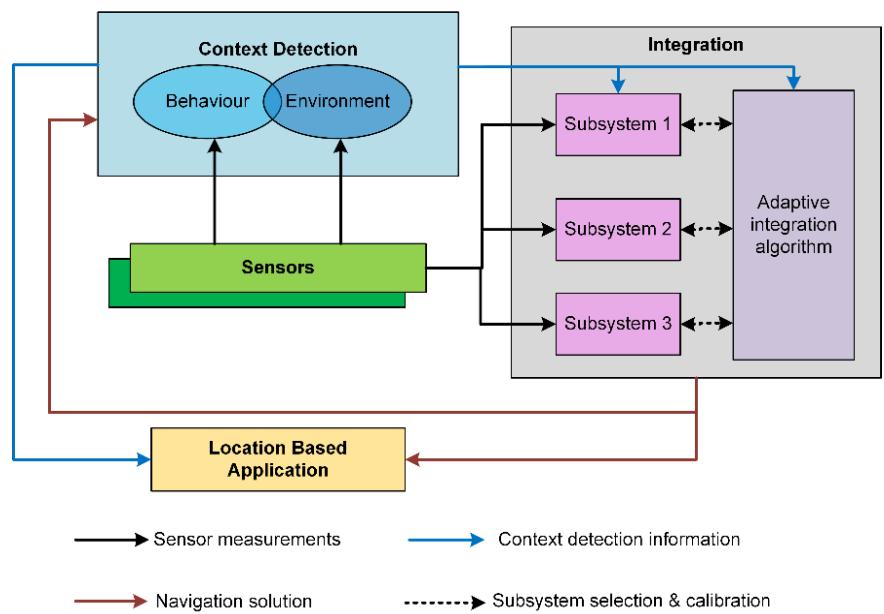

Figure 1. An example of a context adaptive navigation system

Figure 1 illustrates a possible architecture for a multi-sensor context adaptive navigation or positioning system. In a self-aware context adaptive navigation system, behavioural and environmental context categories are identified from available sensor measurements and context association. Based on the contexts detected, different sensors may be selected and their measurements may be processed in different ways within each subsystem. Consequently, the integration module can adapt itself and export positioning results by selecting the most appropriate subsystems and varying the tuning of the algorithms. Then the positioning results can be used for location based services and improvement of context detection. Different hardware and software modules within the system can be shared or re-used by multiple subsystems and applications to take advantage of the available resource.

Previous work on context navigation has focused on individual subsystems. For instance, cognitive GNSS has been investigated to adjust the processing strategies and parameters of GNSS receivers [13][14][15][16]. Awareness of motion type and sensor location has been implemented for PDR [17][18][19]. Due to its popularity and the rich variety of built-in sensors, the smartphone has become an ideal platform for testing and demonstrating contextual experiments. Meanwhile, the rapid advance of machine learning provides the tools to infer context from the smartphone's sensor measurements. There has been substantial research into determining activity recognition for indoor positioning applications [20][21][22]. Using the "IODetector" proposed by Zhou [23], indoor/outdoor environment is determined by using a combination of cellular signals, light sensors, magnetometers and proximity sensors. With the same sensors, semi-supervised learning has also been considered to improve the detection accuracy [24]. Besides, other sensors, such as Bluetooth [25] and microphone [26], have also been utilised for indoor and outdoor detection.

A number of researchers have investigated different approaches towards a context adaptive navigation system. A Location-Motion-Context (LoMoCo) model [27] was proposed using Bayes reasoning to determine the context information from the locations and motion states. At the same time, an activity and environment recognition method with an adaptation algorithm for context model parameters was described in [28]. In 2013, a 'context adaptive navigation' framework was introduced systematically in [1] by UCL, with the preliminary behavioural and environmental context detection results following. Following the initial proof of concept, a basic context detection system was presented in [29], including context categorisation, behaviour recognition, environment detection from GNSS signals and a simple context association demonstration.

Existing work have demonstrated the relevant context detection techniques and built the foundation of a context adaptive navigation system. To further extend this research, this paper aims to show how behavioural and environmental context can be detected and associated for a reliable context determination. It is structured as follows. Section 2 describes context categorization and the framework of context detection algorithm. Section 3 considers behaviour recognition, including the selection of features and classifiers as well as behaviour connectivity. Section 4 proposes the investigation of environment detection, including classification and smoothing. Section 5 investigates context association to connect two aspects of context. Section 6 then presents the performance of context detection algorithms under different scenarios. The conclusions are summarized in Section 7.

2. CONTEXT DETECTION FRAMEWORK. Figure 2 shows the components of the context detection algorithm. Firstly, behaviours are recognised from accelerometers, gyroscopes, magnetometers and the barometer on the smartphone while environments are detected from GNSS signals. Different features are extracted from the sensor measurements and used in machine learning algorithms for classification. Then, the behaviour and environment classification results are smoothed by the proposed behavioural connectivity method and HMM respectively, aiming to minimise incorrect context determination by taking advantage of the relationships between epochs. Within the algorithm, context association is also implemented to further enhance context determination reliability as behaviours and environments are not completely independent in reality. Based on the behaviour recognition results, the parameters in HMM are adjusted with the probabilities of being stationary states. The outputs of both classification and smoothing processes are estimated in probability, so that the navigation system can make decisions according to the certainty of the context detection results.

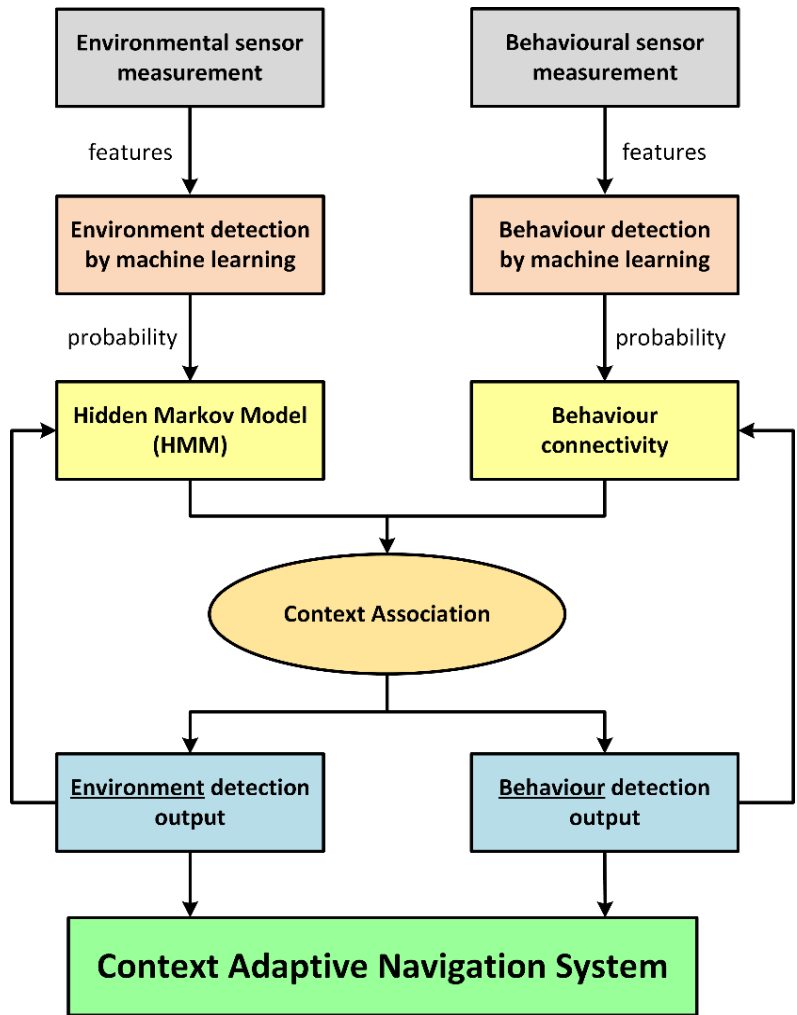
Figure 2. Diagram of context detection algorithm

3. BEHAVIOURAL CONTEXT RECOGNITION. A two-step process is implemented for behavioural context recognition: pattern recognition algorithms are used to assess the probabilities of the behaviour belonging to each category using features from sensor measurements, then a connectivity algorithm is proposed to consider the detected behaviour alongside the behaviour from the previous epoch.

In this section, the categorisation and behavioural recognition framework is introduced in Section 3.1. A detailed description of classification using the machine learning algorithm can be found in Section 3.2. To enhance the recognition reliability, the connectivity algorithm is further presented in Section 3.3. Note that most of the work in this section was published at ION GNSS+ 2016 [29].

3.1. *Categorisation.* The behavioural context may be divided into several broad classes: pedestrian activity, land vehicle, water vehicle, aircraft and spacecraft [1]. Each class contains detailed subdivisions. To fit the purpose of daily smartphone applications, only the human activity and land vehicle classes are included within the scope of the research.

To provide robust and accurate classification of behaviours, a hierarchical detection frame is proposed to proceed from a coarse-grained recognition towards fine-grained subtasks. As shown in Figure 3, three classifiers are consisted in the framework: a human-vehicle classifier, a human activity classifier and a vehicle motion classifier. The top level of classifier is designed to distinguish between the broad classes while the bottom level of classifiers is responsible for recognising the subclasses within each broad class. A human-vehicle classifier is organized at the top level of the system to distinguish between motorised vehicle motions and non-motorised activities. When motorised transport is recognised, the detection system proceeds to the vehicle motion classifier for classification of different vehicle motions. Otherwise, it proceeds to the human activity classifier. Compared with a single classifier dealing with all behavioural scenarios, this hierarchical framework can select different features and pattern recognition algorithms used in different classifiers for better recognition performance.

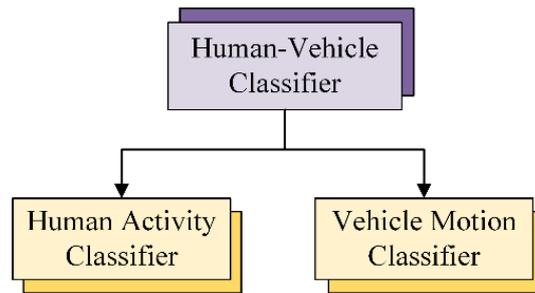

Figure 3. Overview of behaviour recognition framework

A set of detailed categories currently included within each classifier are introduced in Table 1. Note that distinct from human activities, motorised land vehicles, propelled by internal combustion engines or electric motors, sometimes combinations of the two, can be identified by the vibrations from the frequency spectrum of the accelerometers. Engine vibration applies mainly to internal combustion engines, whereas road-induced vibration applies to all land vehicles. The vehicles covered in this study include diesel trains, diesel buses and underground trains. All underground trains are electric for safety reasons. The hybrid vehicles were not included in the current study. The mode of stationary vehicles with the engine on is included within the category because it can play a significant role in context connectivity to minimise impossible behavioural context transitions, such as from a moving vehicle to another moving vehicle directly, or one human activities connected to a moving vehicle without intermediate categories. A further discussion will be presented in Section 3.3.

Table 1. Detailed types of behaviours

| Human activity types | Vehicle motion types |
| --- | --- |
| Stationary; | Stationary vehicles with the engine on; |
| Walking; | Moving diesel trains; |
| Running; | Moving diesel buses; |
| Ascending stairs; | Moving underground trains. |
| Descending stairs. | |

3.2. *Classification Model*. As previous research [30][31] has already proved, among the sensors in a smartphone, measurements from the inertial sensors are capable of taking the leading roles in motion recognition. The accelerometer and gyroscope signals are able to track kinematic motions indirectly by measuring the specific force and angular rate. Motion can also be inferred from some sensors that measure magnetic features. Magnetometers sense the magnetic fields, enabling changes in heading to be detected. A barometer, also called a barometric altimeter, measures the ambient air pressure, from which the heights can be estimated and the changes in height can be derived [2]. Therefore, in this study, accelerometers, gyroscopes, magnetometers and a barometer, found in most smartphones, are used for behavioural recognition.

The construction of a behavioural classification model consists of three main phases: preprocessing, feature extraction and pattern recognition. They are described as follows:

1) Preprocessing: The raw data from smartphone sensors can be processed by various ways, such as removing incorrect data and filtering. Then, a windowing scheme is applied to segment the data into successive pieces for further calculation.

2) Feature extraction: Various features, representing the main characteristics of behaviours, are extracted from the segmented data as the inputs of classifiers.

3) Pattern Recognition: This comprises two stages. In the training stage, the recognition classifiers for classification are constructed and the parameters of the model are learned from training sets. In the classification stage, the trained classifiers are used to recognise different behaviours.

3.2.1. *Sensing and Preprocessing*. Prior to feature extraction, the raw sensor data are divided into small segments using sliding windows. The selection of an appropriate window length is important, and different durations can be set for it. At a sampling frequency of 100 Hz, a 500 sample window is suitable based on previous studies [32][33]. It was shown that a window length of four seconds was an effective and sufficient value for behaviour recognition, neither too short to capture enough features, nor too long to avoid mixing multiple contexts in a single window. A 4s sliding window with a 50% overlap is used for training to avoid missing information between successive windows. Note that to get quicker responses, a 75% overlap is adopted for behavioural context, thus context can be determined every second.

The accelerometers, gyroscopes and magnetometers provide measurements in three dimensions, referred to as the *x*-axis, *y*-axis and *z*-axis. However, the recognition performance may be affected by orientation changes if the model is trained only for a specific orientation [30][34]. In order to minimise such effects, the magnitudes of the sensors are calculated from the outputs of three axes, *x*, *y* and *z*, thus

$$magnitude = \sqrt{x^2 + y^2 + z^2} \tag{1}$$

However, the existence of a sequence with a non-zero mean can hide important information in the frequency domain, so the means of the magnitudes are removed from each segment prior to computing the frequency-domain features.

3.2.2. *Feature Extraction*. Once the data pre-processing is completed, features need to be extracted from the segmented data to be used for training and classification. A good set of feature measurements can often provide accurate and comprehensive descriptions of patterns from which the differences between context categories are easily discerned. In this study, both time-domain and frequency-domain features are extracted for behavioural recognition.

Time-domain features describe temporal variations of motions during the sliding window. The time-domain features selected include range, variance, skewness and kurtosis extracted from all sensors. The effectivenesses of these features for behaviour classification have been shown in different studies [19][30][35]. Zero-crossing rate (ZCR) is also extracted from the preprocessed accelerometer signals, which is used to differentiate different periods of human activity changing with the time. They are expressed as follows and summarized in Table 2.

$$range = max\{x\} - min\{x\} \tag{2}$$

$$\sigma = \sqrt{E\{(x-\mu)^2\}} = \sqrt{\frac{1}{N}\sum_{n=1}^{N}(x_n - \bar{x})^2} \tag{3}$$

$$skewness = \frac{E\{(x-\mu)^3\}}{\sigma^3} = \frac{1}{N\sigma^3}\sum_{n=1}^{N}(x_n - \bar{x})^3 \quad (4)$$

$$kurtosis = \frac{E\{(x-\mu)^4\}}{\sigma^4} = \frac{1}{N\sigma^4}\sum_{n=1}^{N}(x_n - \bar{x})^4 \quad (5)$$

$$ZCR = \frac{1}{N-1}\sum_{n=1}^{N-1}\mathbb{I}\{x_n x_{n+1} < 0\} \quad (6)$$

where $N$ is the number of samples over the window, $\mu$ is the mean, $x_n$ represents the $n$-th epoch of data in the window and the indicator function $\mathbb{I}(.)$ is 1 if its argument is true and 0 otherwise.

Table 2. Behavioural features in time-domain

|  | Expression | Human-Vehicle | Human Classifier | Vehicle Classifier |
|---|---|---|---|---|
| **F1** | $range_{acc}$ | √ | √ | √ |
| **F2** | $range_{gyro}$ | √ | √ | √ |
| **F3** | $range_{magn}$ | √ | √ | √ |
| **F4** | $range_{baro}$ | √ | √ | √ |
| **F5** | $\sigma_{acc}$ | √ | √ | √ |
| **F6** | $\sigma_{gyro}$ | √ | √ | √ |
| **F7** | $\sigma_{magn}$ | √ | √ | √ |
| **F8** | $\sigma_{baro}$ | √ | √ | √ |
| **F9** | $skewness_{acc}$ | √ | √ | √ |
| **F10** | $skewness_{gyro}$ | √ | √ | √ |
| **F11** | $skewness_{magn}$ | √ | √ | √ |
| **F12** | $skewness_{baro}$ | √ | √ | √ |
| **F13** | $kurtosis_{acc}$ | √ | √ | √ |
| **F14** | $kurtosis_{gyro}$ | √ | √ | √ |
| **F15** | $kurtosis_{magn}$ | √ | √ | √ |
| **F16** | $kurtosis_{baro}$ | √ | √ | √ |
| **F17** | $ZCR_{acc}$ |  | √ |  |

Frequency-domain features describe the periodic characteristics of motion during the sample window. In frequency-domain analysis, peaks are centered on different frequency values for different behaviours after a fast Fourier transform (FFT). For this reason, features in the frequency spectrum can reveal significant information on motion periods and vibrations. In the human-vehicle classifier and human activity classifier, the frequency of the largest peak and related spectrum peak magnitude of accelerometers and gyroscopes, are extracted to capture the differences between human and vehicles, and the main temporal periodicity of different human activities. Specifically, according to [1][9], the vehicles always exhibit one or more peaks between 20 Hz and 40 Hz due to vibration and small peaks below 10 Hz when the vehicle is not moving. Thus all frequency domain features of the vehicle classifier are estimated in the following sub-bands instead of the whole spectrum: 0-10 Hz, 10-20 Hz, 20-30 Hz, 30-40 Hz and 40-50 Hz.

The Power Spectral Density (PSD) of signals shows the strength of the energy distributed in the frequency spectrum, thus the PSD of accelerometers is adopted in the vehicle motion classifier to distinguish different vehicle motions with diverse vibrations. For finite time series $x_n$ sampled at a discrete time interval of $\Delta t$ for a total measurement period $T = N\Delta t$, the PSD is defined by

$$S_{xx}(\omega) = \frac{(\Delta t)^2}{T} \left| \sum_{n=1}^{N} x_n e^{-i\omega n} \right|^2 . \tag{7}$$

A summary of the frequency domain features for each classifier is presented in Table 3.

Table 3. Behavioural features in frequency-domain

|  | Expression | Human-Vehicle | Human Classifier | Vehicle Classifier |
|---|---|---|---|---|
| **F18** | *Peak magnitude $_{acc}$* | √ | √ |  |
| **F19** | *Peak magnitude $_{gyro}$* | √ | √ |  |
| **F20** | *Frequency index of F18* | √ | √ |  |
| **F21** | *Frequency index of F19* | √ | √ |  |
| **F22- F26** | *Peak magnitudes of acc in sub-bands: 0-10Hz, 10-20Hz, 20-30 Hz, 30-40Hz, 40-50 Hz* |  |  | √ |
| **F27- F31** | *Peak magnitudes of gyro in sub-bands: 0-10Hz, 10-20Hz, 20-30 Hz, 30-40Hz, 40-50 Hz* |  |  | √ |
| **F32- F36** | *PSD of acc in sub-bands* |  |  | √ |

3.2.3. *Pattern Recognition*. Supervised classification methods learn a model of relationships between the target vectors and the corresponding input vectors consisting of training samples and then use this model to predict target values for the test data [36].

A decision tree algorithm is applied for the human-vehicle classifier with a 98.9% classification accuracy [29]. A relevance vector machine (RVM) whose outputs are in probability, is used for both human activity and vehicle motion classifiers, with 97.6% and 91.0% classification accuracy respectively. The detailed description of the algorithms and comparisons with other supervised machine learning algorithms are discussed in [29]. It also showed that the proposed system achieved an overall 95.1% classification accuracy.

3.3. *Connectivity*. One way of reducing incorrect behaviour determination is to consider the likelihood of behaviour connectivity. Connectivity describes the temporal relationship between the current behaviour category and the previous ones. If a direct transition between two categories can occur, they are connected; otherwise, they are not [1]. For example, stationary vehicle and pedestrian behaviour can be connected directly, whereas moving behaviour of different vehicles is not because a vehicle must normally stop to enable a person to get in or out.

Behavioural connectivity is represented in a probabilistic way. Comparing with Boolean results, there are two advantages. First, a Boolean implementation may occasionally result in the decisions being stuck on incorrect context categories following a faulty selection. This can occur when the correct context category is not directly connected to the incorrectly selected category and the other categories are poor matches to the measurement data. But probabilities are more flexible to increase the directly connected category and minimise the unlikely one. Second, a probabilistic scheme permits the transitions between context categories that are rare, but not impossible.

To illustrate the temporal relationships, the likelihoods of connections between behaviours are listed in Table 4, where the permitted direct connections are set to 0.9 and the unlikely connections are set to 0.1.

Table 4. Behavioural connection matrix ($C$)

(H = human activities, including stationary, walking, running, ascending and descending stairs; V=stationary vehicles with the engine on; U=moving underground trains; T=moving diesel trains; B=moving buses.)

| Current \ Prev | H | V | U | T | B |
|---|---|---|---|---|---|
| **H** | 0.9 | 0.9 | 0.1 | 0.1 | 0.1 |
| **V** | 0.9 | 0.9 | 0.9 | 0.9 | 0.9 |
| **U** | 0.1 | 0.9 | 0.9 | 0.1 | 0.1 |
| **T** | 0.1 | 0.9 | 0.1 | 0.9 | 0.1 |
| **B** | 0.1 | 0.9 | 0.1 | 0.1 | 0.9 |

As the behaviours between two concessive epochs are not independent, a straight smoothing method is first applied. As in Equation (8), the smoothed estimates are obtained by combining the normalised outputs from the classification algorithms at epoch $k$ and the estimates at epoch $k$-1 using filter gain α.

$$\hat{\mathbf{x}}_k^- = \alpha \cdot \mathbf{z}_k + (1-\alpha) \cdot \hat{\mathbf{x}}_{k-1} \tag{8}$$

where $\hat{\mathbf{x}}_k^-$ and $\hat{\mathbf{x}}_{k-1}$ are, respectively, the estimates of behaviours at epoch $k$ before connectivity updating and estimates at epoch $k$-1 and $\mathbf{z}_k$ is the detected probability of behaviours at epoch $k$ across the detection algorithms. α=0.5 is used here, which indicates the measurements at epoch $k$ and the estimates at epoch $k$-1 are weighted equally. Then the relationships between estimates are constructed based on a linear assumption by a transfer matrix $\mathbf{\Omega}_k$, as shown in Equation (9).

$$\hat{\mathbf{x}}_k^- = \mathbf{\Omega}_k \cdot \hat{\mathbf{x}}_{k-1} \tag{9}$$

The transfer matrix is a quantitative representation to describe the response of estimate at epoch $k$ to the previous one. Note that in most practical cases, the dimensions of vector $\hat{\mathbf{x}}_k^-$ and $\hat{\mathbf{x}}_{k-1}$ are larger than one, thus Equation (9) becomes an underdetermined equation. To obtain the transfer matrix, the minimum (Euclidean) norm of the transfer matrix constraint is imposed as it is able to control the propagation to the perturbations in the estimates 21[37][38]. To calculate the matrix, a Moore-Penrose pseudoinverse [39] of vector $\hat{\mathbf{x}}_{k-1}$ is applied:

$$\mathbf{\Omega}_k = \hat{\mathbf{x}}_k^- \cdot (\hat{\mathbf{x}}_{k-1})^\dagger \tag{10}$$

In Equation (10), superscript † is the operator of the pseudoinverse (right inverse in this case), which satisfies

$$\hat{\mathbf{x}}_{k-1} \cdot (\hat{\mathbf{x}}_{k-1})^\dagger = \mathbf{I} \tag{11}$$

However, connectivity implies that some transitions are more likely than others, thus the transfer matrix should be re-estimated using the connectivity constraints, as shown in Equation (12). In Equation (12), notation $\circ$ denotes matrix element-wise multiplication, satisfying $(\mathbf{\Omega} \circ \mathbf{C})_{i,j} = \mathbf{\Omega}_{i,j}\mathbf{C}_{i,j}$.

$$\hat{\mathbf{x}}_k^+ = (\mathbf{\Omega}_k \circ \mathbf{C}) \cdot \hat{\mathbf{x}}_{k-1} \tag{12}$$

The final step is to re-scale the likelihood of each category to obtain a probability using

$$\hat{x}_{k,i} = \frac{\hat{x}_{k,i}^+}{\sum_I \hat{x}_{k,i}^+} \tag{13}$$

where $\hat{x}_{k,i}$ is the probability of behaviour $i$ at epoch $k$.

4. ENVIRONMENT CONTEXT CLASSIFICATION. Different kinds of radio signals are inherently environment-dependent and may be used for environment classification. The section begins by explaining why GNSS signals have been selected for detecting the environments. The extraction of suitable features based on the availability and strength of GNSS signals is then described. Next, the environments are classified by a probabilistic support vector machine (SVM) in contrast to the heuristic approach described in [29]. Finally, the SVM approach is smoothed by a hidden Markov model (HMM) to improve the context determination.

4.1. *Categorisation*. A navigation system may not fully benefit from a context framework that is designed for more general purposes. The context categorization framework for navigation and positioning must be designed especially in order to be fit for its purposes. A good environment categorization for navigation is expected to provide an indication of the positioning techniques applicable for determining position in that environment.

For smartphone applications, a common mobile user spends most of their daily life on land, thus the range of environmental contexts in this study only consider scenarios on land. For land navigation, locating whether the user is indoor or outdoor is a basic but prerequisite task because indoor and outdoor positioning depend on inherently different techniques. For example, in an outdoor environment, GNSS or enhanced GNSS techniques perform well while Wi-Fi positioning or Bluetooth positioning are better options when staying inside a building. In reality, the boundaries between indoor and outdoor environments can be ambiguous, rendering some scenarios hard to classify as either indoor or outdoor [28]. Thus, the intermediate environment, where a client is adjacent to a building or in a partially enclosed environment, is included as one of the categories. Typical examples of intermediate environments are shown in Figure 4, where the areas above are covered by the building and at least one surrounding side of the area is open to the outside. Note that the occurrence of intermediate scenarios is quite rare for vehicles, so it is ignored when constructing the vehicle model.

Table 5. Environment categorisation

| Pedestrian | Vehicle |
|---|---|
| Indoor; Intermediate; Outdoor. | Indoor; Outdoor. |

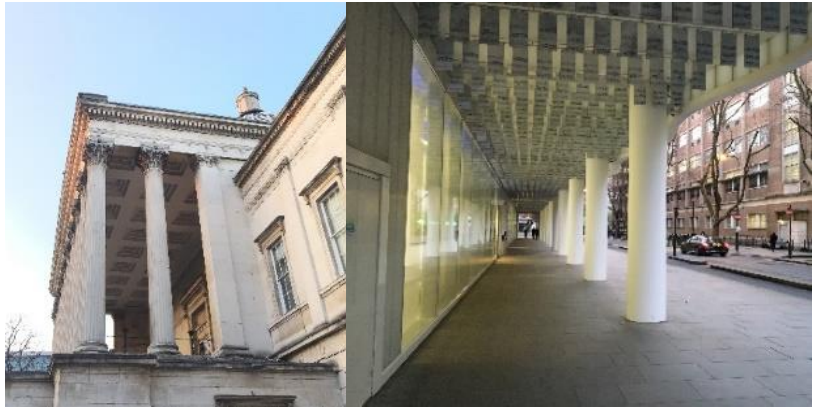

Figure 4. Examples of intermediate environments

Different smartphone sensors whose outputs vary with features of the environment can be potentially used as detectors and each sensor used for environment detection has its respective advantages and drawbacks. A cellular module detects cellular signal strengths from a cellular network, but at the same time the signals strongly depend on the proximity of cellular base stations in the network. A Wi-Fi module can receive signals broadcast from access points. However, tests [1][9] have found that the assumption, the number of access points are larger and strength of signals are stronger indoors, does not always stand. Thus it was not possible to reliably distinguish indoor environments from outdoor environments using Wi-Fi signals alone. A GNSS module, with GPS (Global Positioning System) and GLONASS (GLObal NAvigation Satellite System) chips in most current smartphones, is chosen as the main detector for this research, because the availability and accuracy of satellite signals tend to be less affected by factors other than the environment type. More importantly, the globally distributed properties of GPS and GLONASS ensure that we can infer environments from the availability and strength of GNSS signals anywhere on Earth. Note that Galileo and BeiDou System can also be used and smartphone GNSS chips start processing their signals. The main drawback of GNSS is its high-power consumption compared to other smartphone sensors. As the research advances, other sensors can be added into the context determination framework to improve upon the environment detection using the GNSS module.

4.2. *Feature Extraction*. In an indoor environment, most GNSS signals are attenuated by the structure of the building or received by NLOS paths, rendering them weaker or unavailable indoors compared with intermediate and outdoor environments. Thus features comprising the total number of satellites received and the total measured carrier-power-to-noise-density ratio ($C/N_0$) summed across all satellites received at each epoch are extracted.

To show the effectiveness of the proposed features, a set of GNSS measurements was collected by smartphone over the transition from an outdoor to an indoor environment. The person holding the smartphone walked from an outdoor to an indoor environment at about the 30th second. Figure 5 and Figure 6 demonstrate the differences in availability and strength of GNSS signals, respectively, in the indoor and outdoor environments. In Figure 5, the number of satellites received decreased gradually after moving indoors, as more satellite signals were blocked by the building. $C/N_0$, expressed in decibel-Hertz (dB-Hz), is a good indicator of signal strength in the absence of significant interference and adopted as a standard output of GNSS receivers. Figure 6 shows the $C/N_0$ outputs from three satellites. A drop of about 5 dB-Hz was observed when the person was nearing the building, following by a sharp decrease when they entered. It was also noted that most of the satellite signals indoors were weaker than 20 dB-Hz and PRN 83 lost track after about 90s.

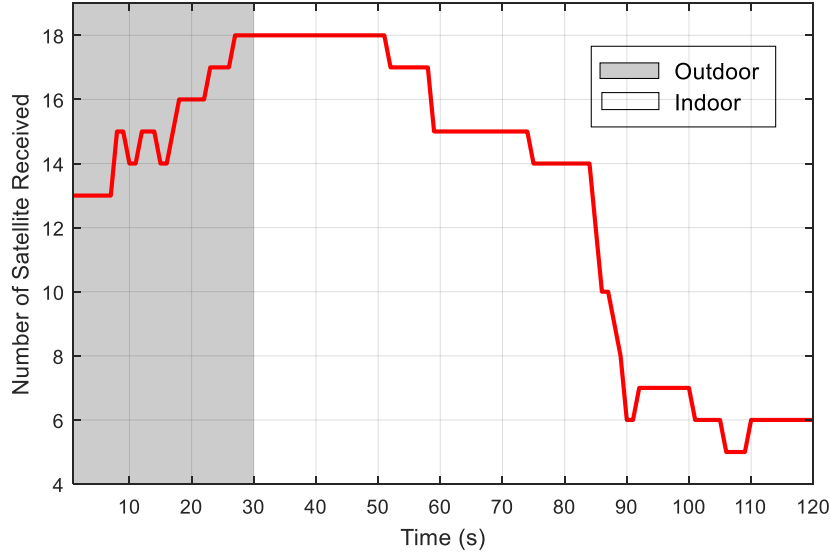

Figure 5. Number of satellites received during outdoor-indoor transition

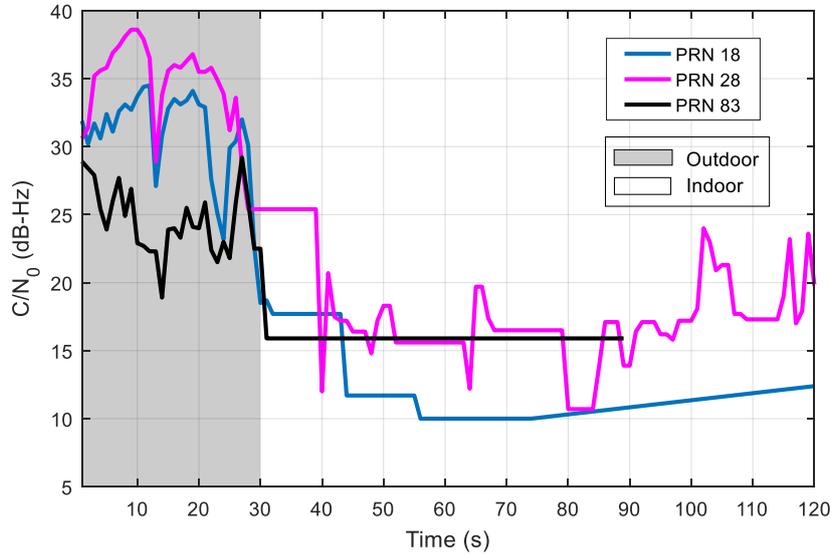

Figure 6. Selected C/N$_0$ values during outdoor-indoor transition

However, based on the findings in [29], it is difficult to distinguish "shallow indoor" and "deep urban" from each other with these two features, so more metrics are required for a reliable indoor/outdoor classification. As proposed in [29], the threshold based metric, total values summed across the satellite signals above 25 dB-Hz, was shown to be effective in indoor/outdoor classification for a pedestrian. Therefore, it is adopted as the third feature for pedestrian based environment classification, denoted as *sumCNR$_{25}$*.

4.3. *Probabilistic Support Vector Machine*. Fundamentally, the SVM is a binary classifier derived from statistical learning theory and kernel based methods [36][40]. In the training phase, given the training samples $X=\{x_i | i=1,2,\cdots,N\}$ with corresponding labels ($y_i \in \{0, 1\}$), the SVM learning classifier is constructed to find the optimum hyperplane in the high-dimensional feature space that maximises the margin between two classes and minimises the error. As shown in Figure 7, the distance of the hyperplane to the nearest training data points of any classes is called the optimal margin and those samples on the margin are called support vectors.

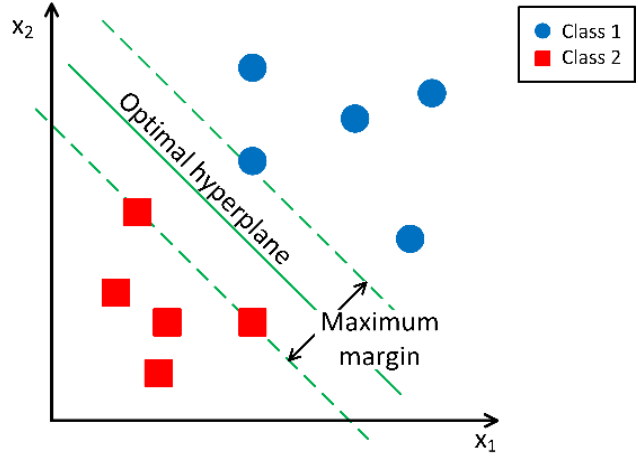

Figure 7. Classification of a non-linearly separable case by SVMs

For a nonlinear classification problem, samples are spread out by being mapped from the original space into a higher dimensional space via the nonlinear transformation function $\Phi(\cdot)$, making the hyperplane easier to be defined in the new space. To reduce the computational load, a kernel function $\kappa$ is defined to substitute the dot products of the transformed vectors.

$$\kappa(\mathbf{x}_i, \mathbf{x}_j) = \Phi(\mathbf{x}_i)^T \Phi(\mathbf{x}_j) \tag{14}$$

Then the hyperplane can be found by solving a constrained optimisation problem:

$$\arg\min_{\mathbf{w},\xi} J(\mathbf{w},\xi) = \arg\min_{\mathbf{w},\xi} (\frac{1}{2}\|\mathbf{w}\|^2 + \beta \sum_{i=1}^{N} \xi_i) \tag{15}$$

subject to the condition:

$$\begin{aligned} (\mathbf{w}^T \Phi(\mathbf{x}_i) + b) y_i &\geq 1 - \xi_i \\ \xi_i &\geq 0, \ i=1,2,\cdots,N \end{aligned} \tag{16}$$

where the hyperplane is defined by the parameters of *w* and *b* as $\mathbf{w}^T\Phi(\mathbf{x}) + b = 0$, $\xi_i$ is the slack variable to tolerate the effect of misclassification of training data, $\beta$ is a positive regularisation parameter specified by the user, determining the trade-off between the training error and the margin size. The above optimisation problem can be solved by the use of Lagrange multipliers, as shown in Equation (17):

$$\begin{aligned} \mathcal{L}(\mathbf{w}, b, \xi, \boldsymbol{\alpha}, \mathbf{r}) \\ = J(\mathbf{w}, \xi) - \sum_{i=1}^{N} \alpha_i (y_i(\mathbf{w}^T \Phi(\mathbf{x}_i) + b) - 1 + \xi_i) - \sum_{i=1}^{N} r_i \xi_i \end{aligned} \tag{17}$$

with $\alpha_i, r_i \in \mathbb{R}$ being the Lagrange multipliers and function $J(w, \xi)$ is defined in Equation (15). Note that the training samples are support vectors if and only if the corresponding multipliers are non-zero. To minimise the above Lagrange function $\mathcal{L}$, we calculate the optimal values of *w*, *b*, $\xi_i$ such that the partial derivatives of $\mathcal{L}$ with respect to these parameters are zero, then the problem becomes to find the equivalent optimisation solution:

$$\max_{\boldsymbol{\alpha}} \left[ \sum_{i=1}^{N} \alpha_i - \frac{1}{2} \sum_{i,j=1}^{N} \alpha_i \alpha_j y_i y_j \kappa(\mathbf{x}_i, \mathbf{x}_j) \right]$$

subject to $\alpha_i \geq 0, \ i=1,2,\cdots,N$ . (18)

$$\sum_{i=1}^{N} \alpha_i y_i = 0$$

After training and finding the $w$, given an unknown sample $x_k$ measured on features, we can classify it by looking at the sign of:

$$f(\mathbf{x}_k) = \sum_{i=1}^{N} \alpha_i y_i \kappa(\mathbf{x}_k, \mathbf{x}_i) + b \ . \tag{19}$$

$f(\mathbf{x})$ is a distance measure between the test sample $x_k$ and the hyperplane defined by the support vectors, thus it cannot be directly used as a probability estimate. Platt [41] proposed an estimate for the posterior class probability by fitting the SVM output with a sigmoid function:

$$P(y_k = +1 | \mathbf{x}_k) = \frac{1}{1 + \exp(Af(\mathbf{x}_k) + B)} \ . \tag{20}$$

The parameters $A$ and $B$ of Equation (20) are found using maximum likelihood estimation from a training set and target probabilities $t_i = (y_i+1)/2$. Note that the training set can be but does not have to be the same set as used for training the SVM [42].

In order to tackle multiclass situations using the SVM method, two possible strategies could be used [36]. The first one is the 'one-against-all' strategy. $L$ binary classifiers will be created for a $L$-class classification and each classifier is trained to separate one class from the others. The second strategy is 'one-versus-one'. There are $L(L-1)/2$ binary classifiers created to separate every two classes. There are $L=3$ environmental contexts in our case, so the two methods have the same computational efficiency. The 'one-versus-one' strategy is adopted.

For each SVM we get using Platt's method, these pairwise probabilities are combined into posterior probabilities by [43]

$$P(S_i | \mathbf{x}_k) = 1 / \left[ \sum_{j=1, j \neq i}^{L} \frac{P(S_i \text{ or } S_j | \mathbf{x}_k)}{P(S_i | \mathbf{x}_k)} - (L-2) \right] \tag{21}$$

where $S_i$ denotes the environment context in this research.

4.4. *Hidden Markov Model.* A hidden Markov model is a temporal pattern recognition algorithm, which assumes a Markov process [30] with the states (indoor, intermediate or outdoor environment in this study), so that it is capable of modelling the movements of a smartphone from one environment to another according to observations. Within an HMM, the probabilities that the system is in each of three states are estimated, so that the navigation system knows the certainty of the decision. In general, an HMM comprises five elements as follows:
1) The state space $S$ that consists of three hidden states: indoor, intermediate and outdoor, which are denoted as $S_1$, $S_2$ and $S_3$ respectively. At each epoch $k$, hidden states satisfy the condition

$$\sum_{i=1}^{3} P(Z_k = S_i) = 1 \tag{22}$$

where $Z_k$ refers to the environmental context at that epoch.
2) The set of observations $x_k$ at each epoch $k$, comprising the extracted features for environment detection.
3) The matrix of state transition probabilities $A=\{A_{ij}\}$. Each element of the state transition probabilities matrix, $A_{ij}$, defines the probability that a state $S_i$ at the immediately prior epoch transits to another state $S_j$ at the current epoch.
4) The matrix of emission probabilities $B=\{B_i(k)\}$ that defines the conditional distributions $P(x_k | S_i)$ of the observations from a specific state.

5) An initial state probability distribution $\pi=\{\pi_i\}$ that defines the probability of state $S_i$ being at the first epoch.

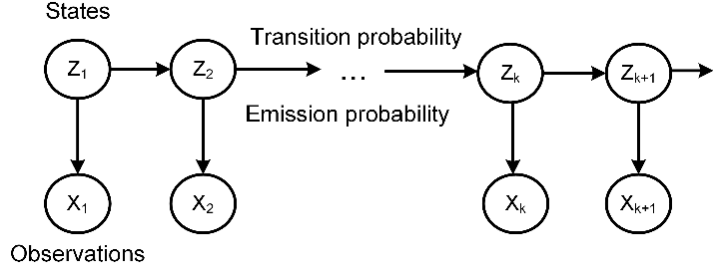

Figure 8. Structure of a first-order HMM

In this paper, we use the first-order HMM, which assumes the current environmental context is only affected by the immediate previous context. Figure 8 is an illustration of a first-order hidden Markov model. Given the sequence of the observations, the probabilities of each context at each epoch can be inferred using the Viterbi algorithm [30][40]. The probabilities of the model are determined as follows.

- Initial probability

As there is no prior information about the initial state, we have to make a judgement based on the available initial observations. Clearly, the indoor and outdoor contexts occur much more frequently than the intermediate context. However, if there is insufficient information to correctly determine the context, it is better to select the intermediate context than to incorrectly select the indoor or outdoor context. The initial probabilities were therefore set as follows:

$$P(Z_1 = S_1) = P(Z_1 = S_3) = 0.4$$
$$P(Z_1 = S_2) = 0.2 \quad (23)$$

- Transition probability

Since the sample interval here is 1s, when a user was previously indoors, the current state is highly likely to be indoor and might be intermediate, but is not likely to be outdoor. This is because the user rarely moves directly from indoors to a fully outdoor GNSS reception environment. However, when the user is at the intermediate state, he/she can move directly to either of the other states. Based on these assumptions and with reference to the parameters applied in IODetector [23], the values of the transition probability were as listed in Table 6. Note that the values are selected in order to obtain an experimental balance between responsiveness to change and vulnerability to noise. There is no intention here to model realistic transition probabilities as this would result in the context determination algorithms taking a long time to respond to changes.

Table 6. Transition probabilities of HMM ($A_0$)

| $k$ \ $k+1$ | Indoor | Intermediate | Outdoor |
|---|---|---|---|
| Indoor | 2/3 | 1/3 | 0 |
| Intermediate | 1/3 | 1/3 | 1/3 |
| Outdoor | 0 | 1/3 | 2/3 |

- Emission probability

The emission probability describes the measurement likelihood of making an observation in different states. Applying the outputs from SVM, emission probability here is transformed from the posterior probabilities in Equation (21) by using Bayes' rule

$$P(\mathbf{x}_k | S_i) \propto \frac{P(S_i | \mathbf{x}_k)}{P(S_i)} \quad (24)$$

where the prior probability of class $P(S_i)$ is estimated by the relative frequency of the class in the training data. In this study, $P(S_1)$=0.4, $P(S_2)$=0.2 and $P(S_3)$=0.4 are used based on two assumptions that indoor/outdoor environments appear more frequently than the intermediate one; indoor and outdoor scenarios have roughly equal appearances.

5. CONTEXT ASSOCIATION. Although behavioural and environmental context are detected separately, they are not completely independent in reality [1]. This information can help the context determination system select a correct context. In this section, behavioural and environmental context association is explored and represented. The transition probabilities in the HMM are tuned differently based on the behaviour recognition results.

The transition matrix $A_0$ given by Table 6 is proposed for general cases without considering the behaviours of the users. In reality, a stationary user will stay in the same environment, making it is impossible to transit from one to another. Therefore, the stationary probability from behaviour recognition results is used to modify the transition probability, as shown in Equation (25).

$$\mathbf{A} = p_{\text{stat}} \times \mathbf{I} + (1 - p_{\text{stat}}) \times \mathbf{A}_0 \quad (25)$$

where $I$ is the identity matrix and $p_{\text{stat}}$ denotes the detected probability of being stationary for both a human and vehicle.

If the user is stationary ($p_{\text{stat}}$=1), the transition matrix will be equal to the identity matrix, indicating an unchanged environment; if the user is detected to be moving ($p_{\text{stat}}$=0), the transition proposed for general cases will be used in HMM.

6. EXPERIMENTS AND DISCUSSION. In this section, different application scenarios were used to test the performance of the proposed context detection system. The performance of static environment detection under different scenarios, kinematic environment detection and the proposed context connectivity method are examined in Section 6.1 and Section 6.2 respectively.

Both behavioural and environmental data was collected using Google Pixel smartphone running Android data logging applications. The behavioural motions were recorded using the 3-axis accelerometers, 3-axis gyroscopes, 3-axis magnetometers and a barometer of the smartphone. GNSS measurements, comprising time tags, PRN (pseudo-random number) of the satellites, the $C/N_0$ measurements, satellite azimuths and elevations can all be logging in files for processing.

6.1. *Case One*. Five different locations were chosen from the test database to carry out environment detection tests. The respective classification results for these scenarios are depicted in Figure 9.

In the case of the open-sky and deep indoor environments, the detection results are very accurate as all samples of these scenarios are successfully detected with almost 100% probability. The shallow indoor scenario is a little challenging for the detector as more LOS signals and some strong reflected signals can be received through the window. It can be observed from the Figure 9 (b) that most samples are classified to indoor correctly but with some intermediate detections occasionally appearing among them. Meanwhile, from the probabilistic outputs from both SVM and HMM, it can be seen that the detection results are much less certain than the deep indoor and open-sky scenarios. A similar behaviour happens for the data collected in an urban area, which can be explained by the fact that some signals are blocked by the surrounding tall buildings and NLOS signals are also received. Comparing the results using SVM alone, many of the misclassified samples are corrected by the HMM, which shows using HMM as a smoother can further improve the performance of environment detection. In the case of the intermediate environment, more signals are blocked by the roof and side walls, but some NLOS signals can still be received from the side without a wall. The decision certainty is thus lower than the other scenarios and some measurements are misclassified.

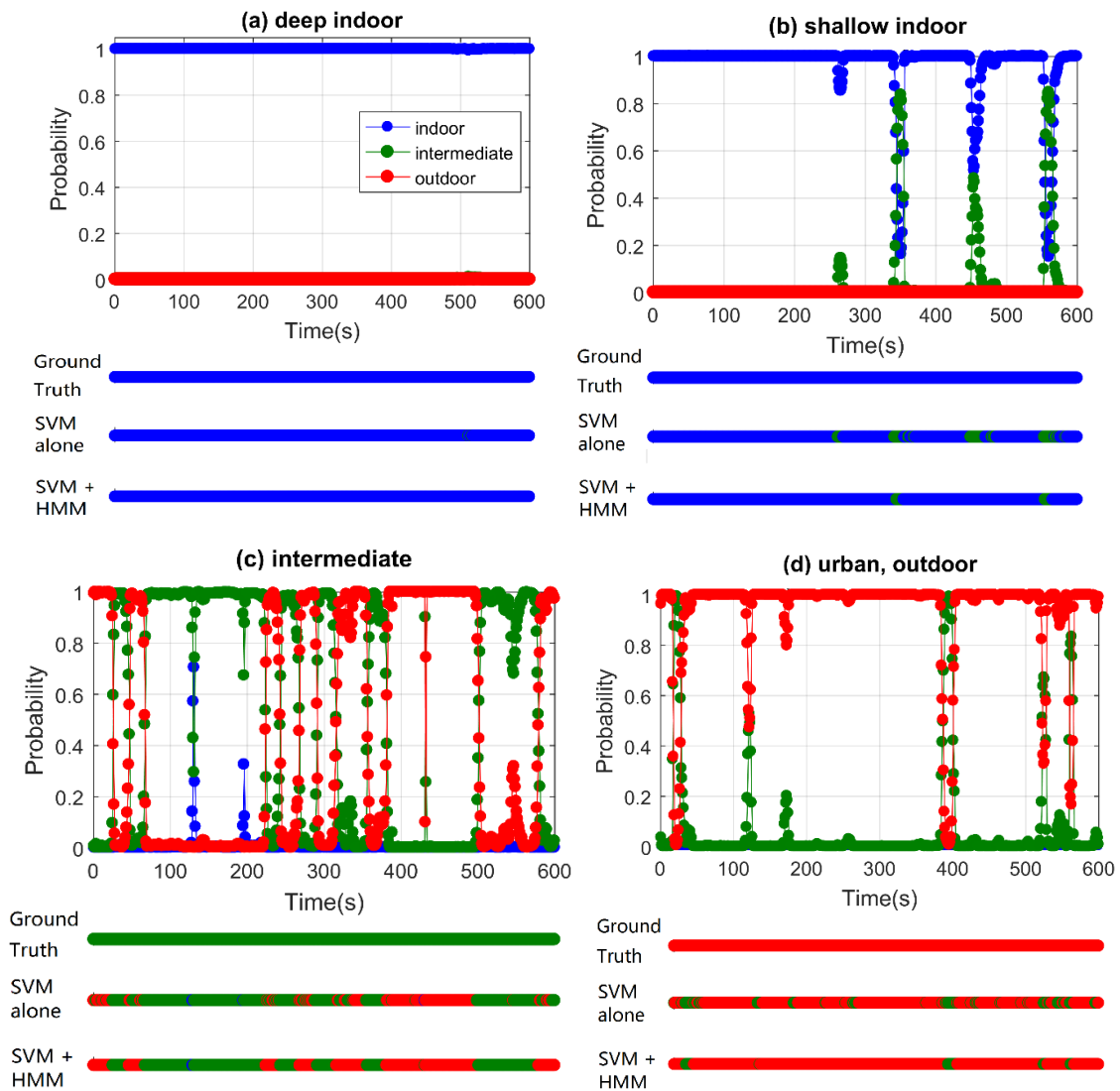

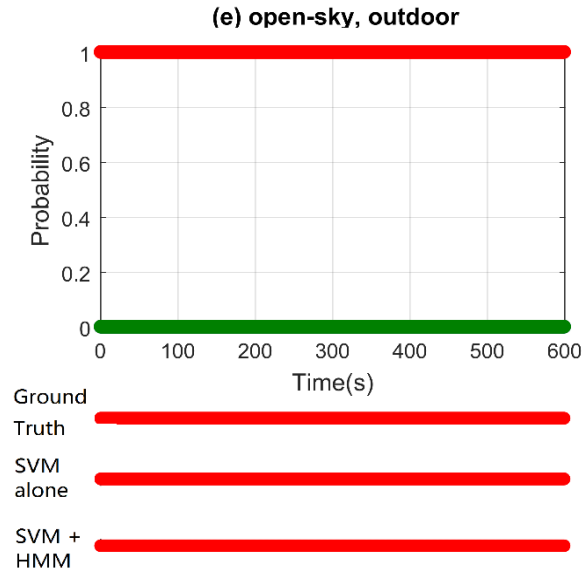

Figure 9. Comparison of static environment detection results under different scenarios with and without HMM

6.2. *Case Two*. To test the performance of the proposed connectivity method, a piece of continuous underground train data was collected on a London underground train (District line) for about 5 minutes, with the vehicle operating and stopping at the stations. It was processed and classified using the same method described in Section 2.2.

A comparison of context recognition results with and without connectivity is shown in Figure 10. Note that most of the misclassified samples are corrected to the right ones, showing that the connectivity constraint is able to reduce the number of incorrect context selections and improve the performance of behavioural detection. Comparing with the reference line, it can also be seen that there were one to two-second response delays after the behaviour changed.

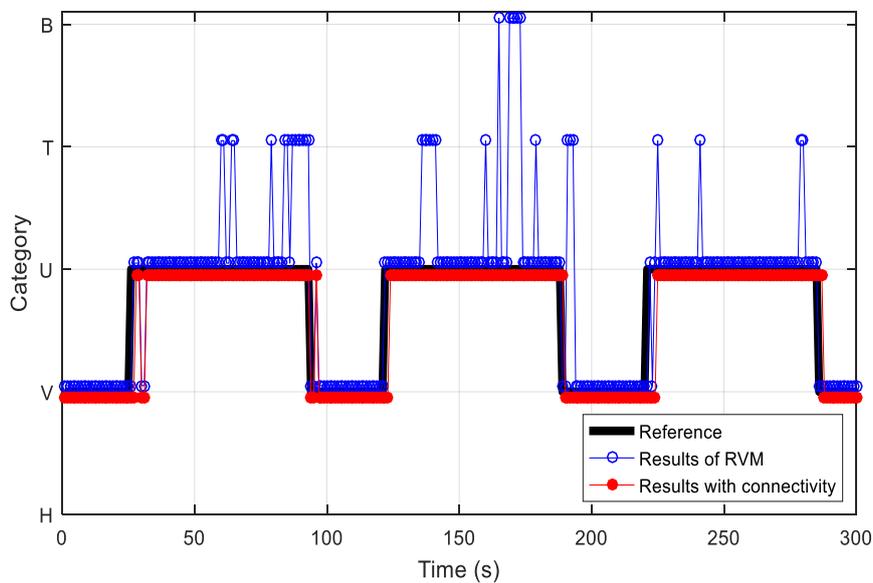

Figure 10. Performance of behaviour detection using connectivity[1]

---

[1] In Figure 10, B=moving buses, T=moving diesel trains, U=moving underground trains, V=stationary vehicles with the engine on, H=human activities.

5. CONCLUSIONS. This paper demonstrates context determination using a smartphone including both the operating environment and the behaviour of a host vehicle or human user, building the foundation of a context-adaptive navigation system.

Detection of behavioural context using accelerometers, gyroscopes, magnetometers and the barometer has been presented. Both time-domain and frequency-domain features are extracted from sensor measurements and classified by supervised machine learning algorithms, which achieved 97.6% accuracy for human activities and 91.0% accuracy for vehicle motions. It has also been shown that the performance can be further improved by considering behavioural connectivity.

Environmental context detection has focused on indoor and outdoor classification. A detection scheme is developed based on GNSS signals and estimating probability of each context. Features of the satellite signals are extracted and classified by SVM. Then a hidden Markov model is used to smooth the results. As behaviours and environments are not independent, context association is applied by using results of behaviour detection to update the transition probabilities within HMM to improve the performance of environment detection.

ACKNOWLEDGEMENTS

This work is funded by the UCL Engineering Faculty Scholarship Scheme and the Chinese Scholarship Council (CSC).